\documentclass[aps,prl,ams,amsmath,twocolumn,reprint,superscriptaddress,longbibliography]{revtex4-2}
\usepackage{blindtext}
\usepackage{graphicx}
\usepackage{graphics}
\usepackage{epsfig}
\usepackage{amsmath}
\usepackage{amssymb}
\usepackage{amsfonts}
\usepackage{color}
\usepackage{bbm}
\usepackage{hyperref}
\usepackage[capitalise]{cleveref}
\usepackage{dcolumn}
\usepackage{bm}

\begin{document}
\title{Quasi-1D Coulomb drag between spin-polarized quantum wires}

\author{Mingyang Zheng}
    \affiliation{Department of Physics, University of Florida, Gainesville, FL 32611, USA}
\author{Rebika Makaju}
    \affiliation{Department of Physics, University of Florida, Gainesville, FL 32611, USA}
\author{Rasul Gazizulin}
    \affiliation{Department of Physics, University of Florida, Gainesville, FL 32611, USA}
    \affiliation{National High Magnetic Field Laboratory High B/T Facility, University of Florida, Gainesville, FL 32611, USA}
\author{Alex Levchenko}
    \affiliation{Department of Physics, University of Wisconsin–Madison, Madison, WI 53706, USA}
\author{Sadhvikas J. Addamane}
    \affiliation{Center for Integrated Nanotechnologies, Sandia National Laboratories, Albuquerque, NM 87185, USA}
\author{Dominique Laroche}
    \altaffiliation{\textbf{Email of Author to whom correspondence should be addressed:} dlaroc10@ufl.edu}
    \affiliation{Department of Physics, University of Florida, Gainesville, FL 32611, USA}

\begin{abstract}

One-dimensional (1D) quantum wires provide a versatile platform for studying strong electron-electron interactions and collective excitations under confinement. Coulomb drag between 1D systems offers a powerful probe of Tomonaga-Luttinger liquid (TLL) physics, with theoretical predictions suggesting distinct power-law in temperature dependencies between the spin-full and the spin-polarized regimes. However, experimental verification has thus far remained limited. Here, we report measurements of reciprocal and nonreciprocal Coulomb drag between vertically coupled quasi-1D quantum wires in the spin-polarized regime. Clear signatures of spin splitting are observed in both the wires conductance and the drag signal. We observed a connection between electron-hole asymmetry and negative drag, and demonstrated different power-law behaviors in spin-full and spin-polarized regimes—yielding consistent TLL interaction parameters. These results validate the theoretical predictions for backscattering induced drag in the reciprocal regime and extend them to the nonreciprocal and the multiple subband regimes. Furthermore, the nonmonotonic density dependence of the reciprocal interaction parameter correlates with the subband occupation of the drag wire, revealing the complexity of the scattering mechanisms in multichannel systems.

\end{abstract}

\maketitle


Strongly correlated electronic phenomena such as correlated insulators \cite{cao_correlated_2018}, electron crystallization \cite{goldman_evidence_1990}, and Bose-Einstein condensations \cite{eisenstein_boseeinstein_2004}, underscore the central role of electron-electron interactions across diverse systems. While two-dimensional (2D) systems have uncovered a range of interaction-driven phases \cite{cao_correlated_2018, cao_unconventional_2018, jiang_controlling_2018, cai_signatures_2023, lu_fractional_2024}, one-dimensional (1D) quantum wires provide an additional controlled and analytically tractable platform for probing such electronic correlations \cite{voit_one-dimensional_1995, DEPICCIOTTO1998395, gogolin2004bosonization}. In 1D, the restricted phase space for scattering and ineffective screening enhance interaction effects, leading to the breakdown of Fermi liquid theory and the emergence of non-Fermi-liquid behavior \cite{haldane_effective_1981, voit_one-dimensional_1995}. Instead of the quasiparticle description in Fermi liquid, the low-energy excitations in 1D systems are propagating collective modes of spin and charge density, described by Tomonaga-Luttinger liquid (TLL) theory \cite{Tomonaga_1950, luttinger_exactly_1963, haldane_effective_1981}.

\begin{figure*}[hbt!]
\includegraphics[width=0.6\textwidth]{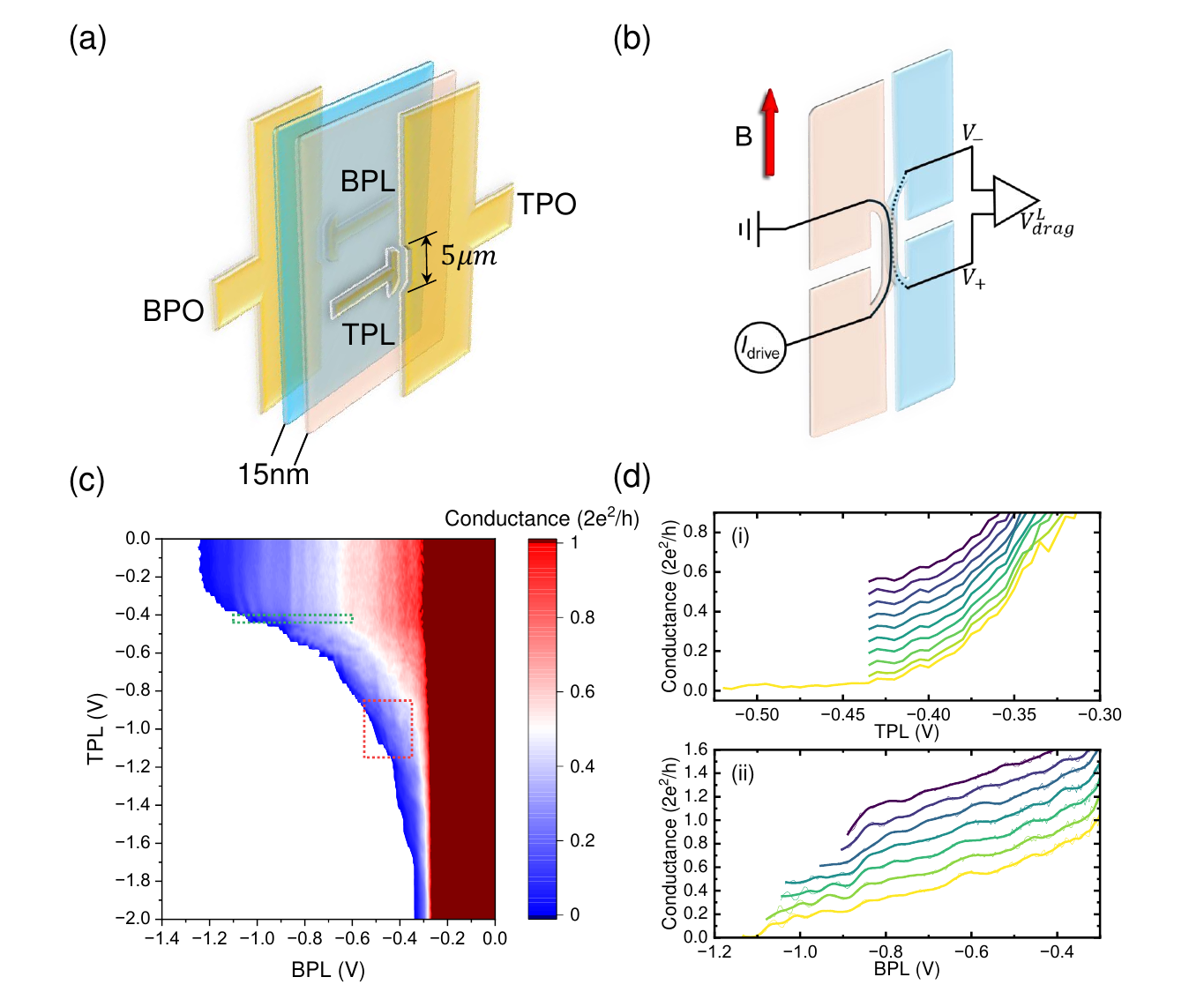}
\caption{\label{fig:1} (a) Schematic of the active part of the double quantum wire device. Each wire consists of a plunger (PL) and a pinch-off (PO) gate. (b) In the interacting region of the device, two vertically superimposed independent quantum wires are created, leveraging selective layer depletion with the PO gates. The top (pink) and bottom (blue) layers are depleted to form the drive and drag wires, respectively. The magnetic field is parallel to both wires, with the direction pointing up as shown in the figure. (c) Bottom (drag) wire conductance 2D map. The green dotted box and red dotted box represent scan window 1 and scan window 2, respectively. 1D subband plateau features are clear and depend mostly on the BPL gate voltages for $TPL > -0.5 \text{ V}$. Interwire crosstalk becomes important and leads to the tilted subband stripes when $TPL < -0.5 \text{ V}$. The top (drive) wire conductance 2D map is similar and is shown in Fig. S4 of the supplement. 
(d) Top (drive) wire (i) and bottom (drag) wire (ii) conductance as a function of bottom and top gate voltages in scan window 1. Plateau-like features are clearly visible in the smoothed data (thick line). (i) The line cuts correspond to BPL = -0.74 V to BPL = -1.1 V from top to bottom, and successive line cuts are vertically offset by 0.03 $2e^2/h$ for visibility. (ii) The line cuts correspond to TPL = -0.38 V to TPL = -0.50 V from top to bottom, and successive line cuts are vertically offset by 0.15 $2e^2/h$ for visibility. The conductance plateaus of both wires are not quantized at integer values of 2e$^2$/h as the wires are non-ballistic. 
}
\end{figure*}

Coulomb drag provides a sensitive probe of interaction and correlation between two coupled low-dimensional systems \cite{rojo_electron-drag_1999}. In a typical drag measurement, current in one wire (the drive wire) induces a voltage response in a nearby, electrically isolated wire (the drag wire), primarily via momentum transfer between charge carriers \cite{narozhny_coulomb_2016}, as shown in Fig. \ref{fig:1}(b). Traditionally, Coulomb drag has been explained by the momentum transfer between charge carriers in the drive and drag wires, which is expected to follow the Onsager relations \cite{Pogrebinskii_1977, gramila_mutual_1991, narozhny_coulomb_2016, Onsager_1931}, yielding a reciprocal signal. Another physical interpretation for the phenomena is current rectification, which interprets Coulomb drag as a rectification of energy fluctuation in the drive circuit \cite{Kamenev_1995, Levchenko_2008, borin_coulomb_2019}. Compared to the reciprocal signal predicted in the momentum transfer model, the current rectification model considers the fluctuating electrical fields and the translational symmetry breaking in mesoscopic circuits, which can lead to a nonreciprocal drag signal whose polarity is determined by the microscopic details of the defect potential in mesoscopic wires in asymmetric systems. Both interpretations yield identical reciprocal drag results in the linear response regime of clean non-correlated systems where electron-hole asymmetry prevents the cancellation of electron and hole contributions \cite{Kamenev_1995}. However, in strongly correlated systems such as TLLs, both approaches produce different predictions \cite{klesse_coulomb_2000, pustilnik_coulomb_2003, fiete_coulomb_2006, dmitriev_coulomb_2012, nazarov_current_1998, Levchenko_2008} and, to the best of our knowledge, no theoretical approach has explicitely combined strong electron-electron interactions with mesoscopic fluctuations.

Experimental studies have confirmed several hallmark signatures of Tomonaga-Luttinger liquid (TLL) physics in 1D systems, including spin-charge separation, power-law scaling, and the upturn of Coulomb drag resistance at low temperatures \cite{bockrath_luttinger-liquid_1999, auslaender_spin-charge_2005, laroche_positive_2011, laroche_1d-1d_2014, wang_one-dimensional_2022}. In parallel, the observation of the 0.7 anomaly in quantum point contacts and quantum wires underscores the richness of electron-electron interactions in 1D systems, particularly in the spin-polarized regime \cite{thomas_possible_1996, thomas_interaction_1998, bauer_microscopic_2013}. While theoretical work has predicted distinctive behavior in Coulomb drag between spin-polarized wires \cite{klesse_coulomb_2000}, experimental exploration of this regime remains largely uncharted. Given the enhanced role of interactions in spin-polarized 1D systems, investigating Coulomb drag in this regime offers a promising avenue to uncover new many-body phenomena.

For 1D systems, Klesse and Stern have predicted different temperature dependencies for Coulomb drag in the spin-polarized and the spin-full regimes \cite{klesse_coulomb_2000}. Below a crossover temperature $T^*$, the electrons in both wires form two interlocked charge density waves (CDWs), resulting in a diverging drag resistance as temperature approaches zero: $R_D\sim R_0 e^ {E_S/T}$, with $E_S \sim T^*$. Above $T^*$, the interlocked CDWs picture breaks down, and $2k_F$ backscattering of the drag wire's electrons by density fluctuations in the drive wire becomes the dominant process. In this case, the Coulomb drag resistance exhibits a power-law temperature dependence, $R_D = R_0 \lambda^2 (\frac{T}{E_0})^x$, with the power $x=4K_{c-}-3$ for spin-polarized wires and $x=2K_{c-}-1$ for spin-full wires. Here, $R_0$ is of order $hk_F/e^2$, $\lambda$ denotes the dimensionless interwire backscattering potential, and $E_0$ is of the order of Fermi energy. The parameter $K_{c-}$ is the TLL parameter of the relative charge-density sector (antisymmetric density mode). Furthermore, Klesse and Stern proposed experiments with the magnetic field parallel to the wires to polarize the electron's spins without affecting their orbital motion.

In this Letter, we report measurements of quasi-1D Coulomb drag between spin-polarized quantum wires. The system consists of two vertically superimposed quantum wires defined electrostatically on a bilayer GaAs/AlGaAs quantum well heterostructure, separated by a thin 15 nm AlGaAs barrier. Each wire is formed within individual two-dimensional electron gases (2DEGs) using surface gates, allowing independent control of their carrier densities and enabling the study of Tomonaga-Luttinger liquid (TLL) interactions in the multi-subband regime. A magnetic field is applied parallel to the wires to achieve spin polarization without introducing orbital effects. By measuring the density and temperature dependence of the drag response, we probe the interaction mechanisms between TLLs in both the spin-full and the spin-polarized regimes. This work presents the first experimental investigation of quasi-1D Coulomb drag in the spin-polarized limit and establishes a platform for studying spin-polarized electron-electron interactions.

\begin{figure*}
\includegraphics[width=1\textwidth]{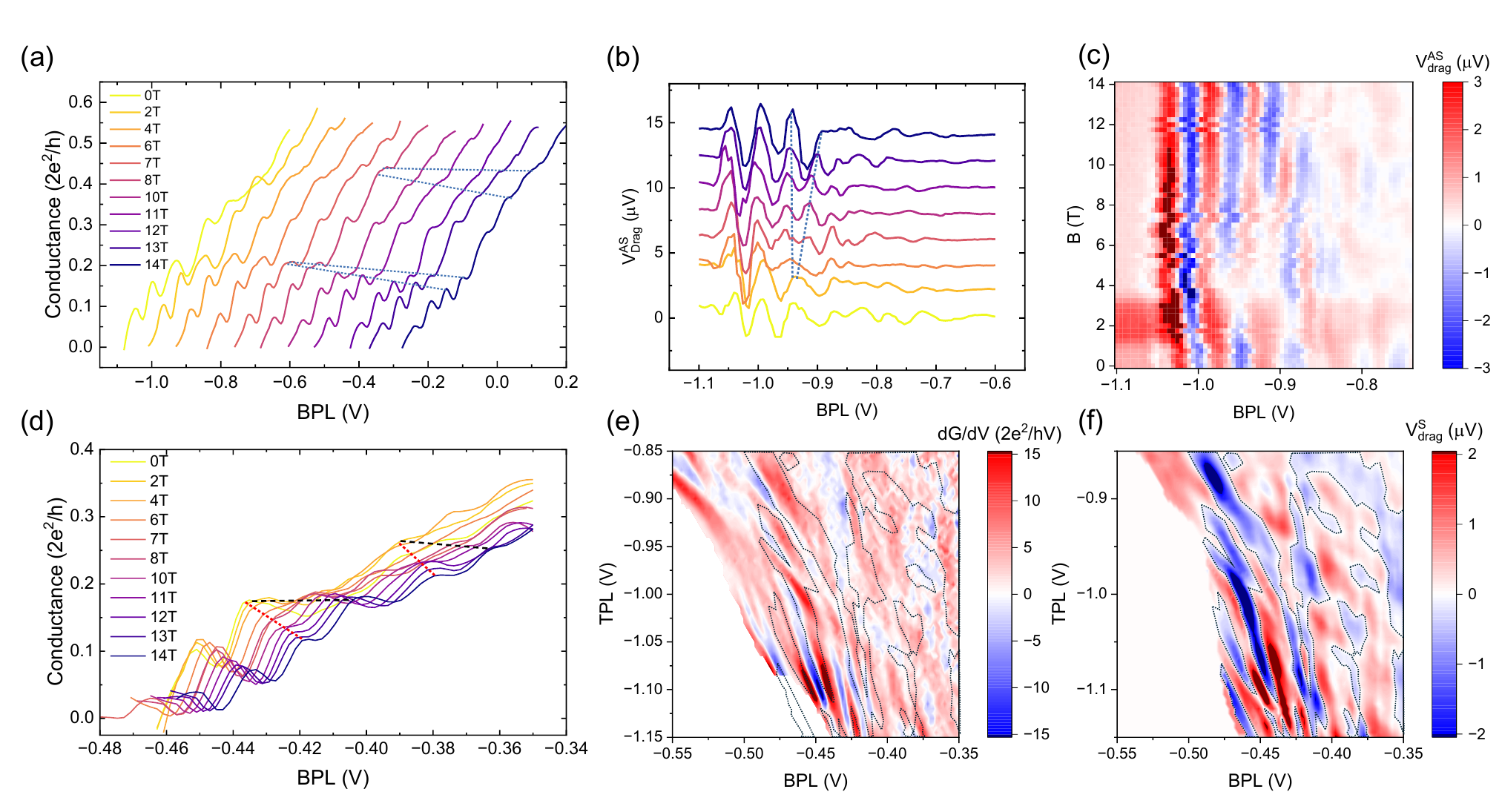}
\caption{\label{fig:2} (a) Drag wire conductance as a function of BPL gate voltage at magnetic fields ranging from 0T to 14 T when TPL = -0.402 V in scan window 1. Successive lines are horizontally offset by 0.08 V for visibility. The 1D subband plateau features and their spin splitting are clearly seen as the magnetic field increases. The dotted lines are guides to the eye of the spin splittings of the first and second subbands. (b) The antisymmetric component of the drag voltage as a function of BPL gate voltage at magnetic fields ranging from 0 T to 14 T from bottom to top, when TPL = -0.402 V. Successive lines are vertically offset by $2 \mu V$ for visibility. The spin-splitting peaks are seen for the first subband around BPL = -0.94 V. The left two peaks around BPL = -1.04 V and BPL = -0.99 V are attributed to defects. (c) The detailed 2D plot of the antisymmetric component of the drag signal as a function of BPL gate voltage and magnetic field for a line cut taken at TPL = -0.402V. (d) Drag wire conductance as a function of BPL gate voltage at magnetic fields ranging from 0T to 14 T when TPL = -1.095 V in scan window 2. The red dotted lines and black dashed lines are guides for the spin-up and spin-down splitting, respectively. (e) The 2D plot of drag wire transconductance as a function of TPL and BPL gate voltages in the scan window 2 at 0 T. (f) The 2D plot of $V_{drag}^S$ as a function of TPL and BPL gate voltages in scan window 2 at 0 T. The dotted lines are the boundaries of the negative $V_{drag}^S$ blocks which are correlated with the drag wire transconductance sign changes in (e). 
}

\end{figure*}

A schematic of the device is shown in Fig. \ref{fig:1}(a) and Fig. \ref{fig:1}(b), and additional details on device fabrication are provided in Appendix A. As shown in Fig. \ref{fig:1}(c) and Fig. \ref{fig:1}(d), both the top and bottom wires exhibit clear 1D conductance plateaus at non-integer values of $2e^{2}/h$, indicative of subband formation in the non-ballistic regime. Owing to its stable subband positions, the top wire is used as the drive wire in the following measurements. By tuning the BPL and TPL gate voltages, the drag signal is measured across various subband configurations in two scan windows defined in Fig. 1. 

\begin{figure*}
\includegraphics[width=1\textwidth]{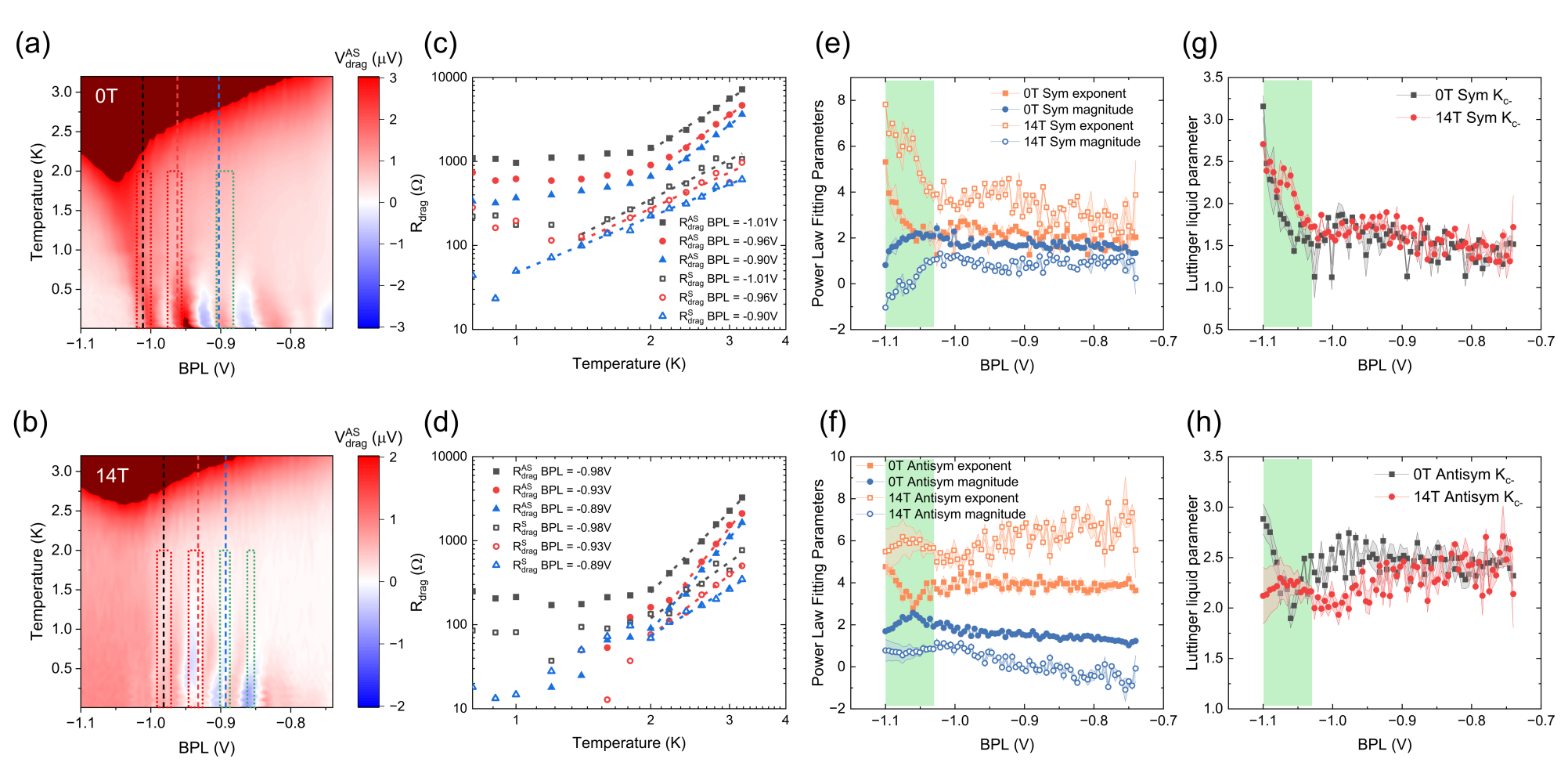}
\caption{\label{fig:3} (a, b) The antisymmetric component of the Coulomb drag voltage, $V^{AS}_{drag}$, as a function of BPL gate voltge and temperature at (a) 0 T and (b) 14 T for a line cut at TPL = -0.424 V in scan window 1. The red dotted boxes represent the first two defects' positions, while the green dotted boxes represent the first subband at 0 T and the first subband spin up and spin down splitting at 14 T. (c, d) Power-law fitting of three different BPL gate positions, as shown by the dashed lines in (a) and (b), respectively. The full symbols represent the $R^{AS}_{drag}$, and the open symbols represent the $R^{S}_{drag}$. The power-law fittings are done in the high-temperature linear range, as shown by the dotted lines. (e, f) Power-law fitting results of the (e) symmetric component and (f) antisymmetric component with the 0T and 14T data shown by the full symbols and observed for both the symmetric and the antisymmetric components. (g, h) The extracted $K_{c-}$ for the two components at 14 T and 0 T. While the exponents are different between 0T and 14T, the $K_{c-}$s have similar values, in agreement with theory \cite{klesse_coulomb_2000}. The green shaded boxes mark the regions where tunneling is significant, and this regime is not further analyzed in this letter.
}
\end{figure*}

Conductance measurements of the drag wire were performed in magnetic fields ranging from 0 T to 14 T. As shown in Fig. \ref{fig:2}(a) and Fig. \ref{fig:2}(d), spin splitting of the first and second subbands is observed with increasing magnetic field at two TPL gate positions, accompanied by a few defect-induced resonances who do not split under magnetic field. Consistent with previous reports, each spin-degenerate 1D subband N splits into $N\uparrow$ and $N\downarrow$, with the $N\uparrow$ conductance gradually decreasing under increasing magnetic field until it reaches one-half of the original spin-degenerate value \cite{Thomas_1996, graham_interaction_2003}. Besides, the $N\downarrow$ plateaus vanish around 10 T and reappear at higher fields near 12 T, consistent with prior studies \cite{graham_interaction_2003} and shown in Fig. \ref{fig:2}(a). This transition is attributed to the Zeeman energy $g\mu_BB_\parallel$ intersecting the subband energy spacing of around 1.01 meV at 10 T.
Corresponding features also appear in the magnetic-field-dependent Coulomb drag signal, as shown in Fig. \ref{fig:2}(b) and Fig. \ref{fig:2}(c). Previous works have revealed significant contributions of both reciprocal and nonreciprocal drag signals in such devices \cite{zheng_tunable_2024}, each showing signatures of TLL physics. To investigate the two contributions in the spin-polarized regime, the drag signal is decomposed into a symmetric component, $V_{\text{drag}}^{S} = \frac{V_{\text{drag}}^R + V_{\text{drag}}^L}{2}$, and an antisymmetric component $V_{\text{drag}}^{AS} = \frac{V_{\text{drag}}^R - V_{\text{drag}}^L}{2}$ \cite{zheng_tunable_2024, zheng_quasi-1d_2024}. The magnetic field dependence of $V_{\text{drag}}^{AS}$ is extracted for a line cut at TPL = -0.402 V and is shown in Fig. \ref{fig:2}(c). $V_{\text{drag}}^{S}$ is shown in Fig. S6 of the supplement and exhibits a similar behavior. The alternating sign of $V_{drag}^{AS}$ aligns with the conductance plateau positions of the drag wire, consistent with previous work \cite{zheng_tunable_2024}. Although negative drag signals have been observed in various systems \cite{anderson_coulomb_2021, du_coulomb_2021, zheng_tunable_2024}, their origin remains an active topic of investigation. Levchenko and Kamenev theoretically demonstrated that such signals may arise from charge rectification driven by electron-hole asymmetry and translation symmetry breaking in mesoscopic circuits \cite{Levchenko_2008}. When the transmission probability for holes exceeds that for electrons at a given energy, a negative drag voltage may result. To investigate this mechanism, we compare the drag wire transconductance and the symmetric drag component in scan window 2, shown in Fig. \ref{fig:2}(e) and Fig. \ref{fig:2}(f), respectively. Due to interwire crosstalk, the positions of the drag wire subbands depend on both the TPL and BPL gate positions, as shown in Fig. S2, producing tilted stripe features in the 2D plots. In mesoscopic 1D wires, imperfections such as boundary roughness or quantum dot–like islands can deform flat conductance plateaus into peak-like structures \cite{nikolic_conductance_1994}. According to the Landauer formula, a decrease in conductance near the right side of a peak indicates a preferred transmission of holes compared to electrons. Thus, negative transconductance is expected to coincide with a positive drag voltage, corresponding to negative drag resistance $R_D = -\frac{V_D}{I_{drive}}$. As shown in Fig. \ref{fig:2}(e) and Fig. \ref{fig:2}(f), we observe a clear correspondence between the sign change of the drag voltage and that of the drag wire transconductance, supporting the interpretation based on electron-hole asymmetry.

\begin{figure*}
\includegraphics[width=1\textwidth]{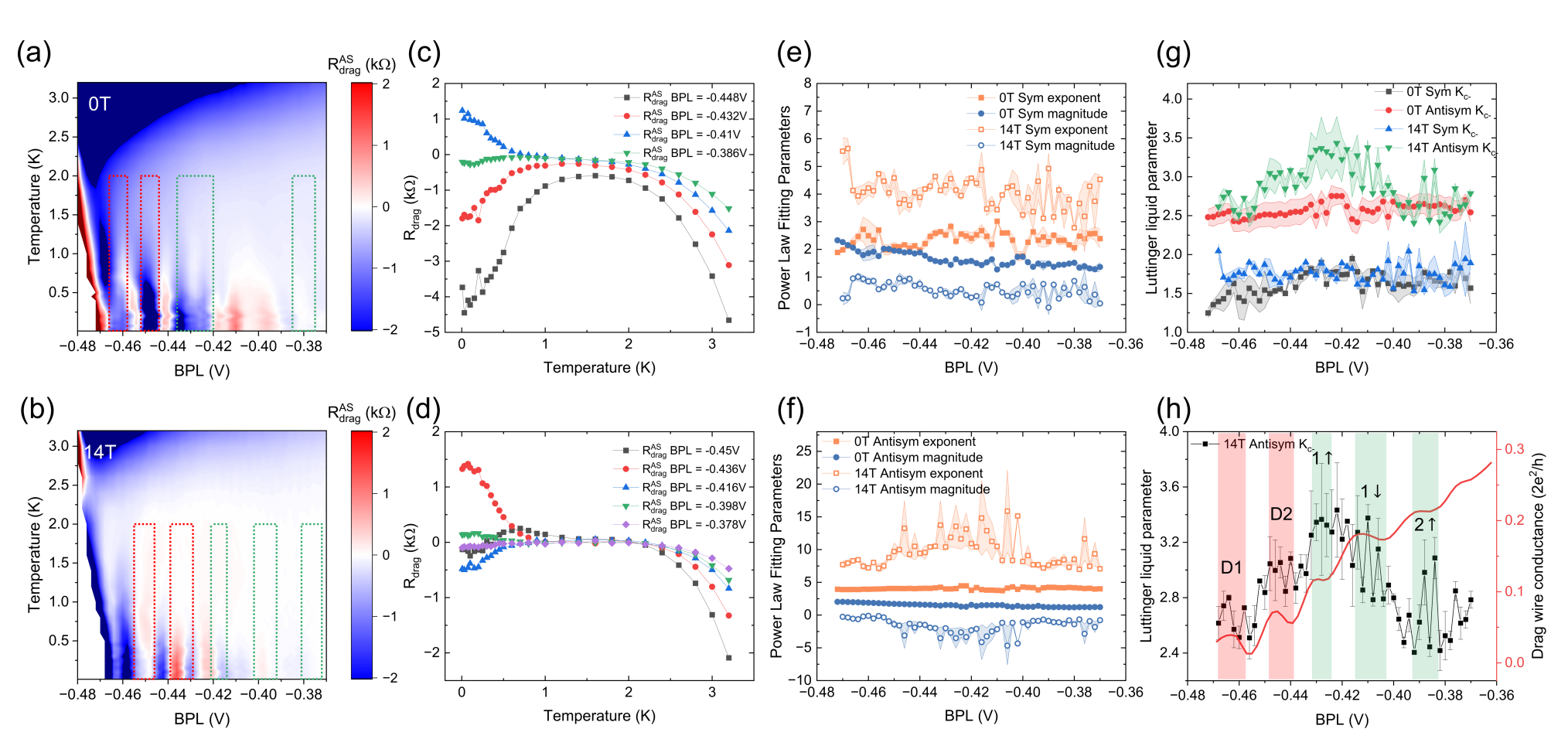}
\caption{\label{fig:4} (a, b) The antisymmetric component of the Coulomb drag voltage $V^{AS}_{drag}$ as a function of BPL gate voltage and temperature at (a) 0 T and (b) 14 T at line cut TPL = -1.095 V in scan window 2. The red dotted boxes represent the first two defects' positions, while the green dotted boxes represent the first and the second subbands at 0T and the first subband spin up, the first subband spin down, and the second subband spin up splittings at 14T from left to right. (c, d) The temperature-dependent $R^{AS}_{drag}$ at several selected BPL gate positions in the dotted boxes of (a) and (b). $R^{AS}_{drag}$ at 0T is shown in (c) and $R^{AS}_{drag}$ at 14 T is shown in (d). (e, f) Power-law fitting results of the (e) symmetric component and (f) antisymmetric component at TPL = -1.095 V with the 0 T and 14 T data shown by the full symbols and open symbols, respectively. The higher exponents and lower intercepts in 14 T are shown for both symmetric and antisymmetric components. (g) The extracted $K_{c-}$ for the two components at 0 T and 14 T. (h) The extracted $K_{c-}$ for the antisymmetric component at 0 T and 14 T and the drag wire conductance at 14 T as a function of the BPL gate positions. Same as (b), the red shades represent the two defect plateaus, and the green shades represent the first subband spin up, the first subband spin down, and the second subband spin up plateau from left to right, respectively. 
}
\end{figure*}

To study the distinct power-law dependence of the drag signal in the spin-full and spin-polarized regimes, we performed temperature-dependent measurements at 0 T and 14 T. As shown in Fig. \ref{fig:3}(a) and Fig. \ref{fig:3}(b), $V_{drag}^{AS}$ first decreases with decreasing temperature until a turning temperature $T_0 \sim 1.5\text{ K}$, below which it increases as the temperature approaches zero. This nonmonotonic behavior is consistent with previous observations and is a key signature of TLL physics \cite{klesse_coulomb_2000, pustilnik_coulomb_2003, laroche_positive_2011, laroche_1d-1d_2014, makaju_nonreciprocal_2024, zheng_tunable_2024}. For $T<T_0$, the drag signal exhibits stripe patterns with alternating signs, which align with conductance plateaus in the drag wire, in agreement with prior results \cite{zheng_tunable_2024}. The subband splitting is also verified through both the drag signal stripes in Fig. \ref{fig:3} and the drag wire conductance measurement in Fig. S10 of the supplement in scan window 1. The temperature dependence of $V_{drag}^{AS}$ and $V_{drag}^{S}$ at selected gate positions is shown in log-log plots with power-law fitting for 0 T and 14 T in Fig. \ref{fig:3}(c) and Fig. \ref{fig:3}(d), respectively. Both components can be fitted linearly in the high-temperature regime, indicating power-law behavior, with $V_{drag}^{AS}$ displaying larger magnitudes and steeper slopes than $V_{drag}^{S}$. In the spin-polarized regime, both components show a reduction in magnitude and an increase in slope, reflecting a change in the power-law exponents. These effects are more clearly seen by directly comparing the fitting results at 0 T and 14 T in Fig. \ref{fig:3}(e) and Fig. \ref{fig:3}(f) for $V_{drag}^{S}$ and $V_{drag}^{AS}$, respectively. It is worth noting that the rise of the power-law exponent and the decrease of the magnitude near the left end of Fig. \ref{fig:3}(e) and Fig. \ref{fig:3}(f) are observed in a regime dominated by tunneling, and is not further analyzed in this letter. In the spin-full regime, the power-law exponents for $V_{drag}^{S}$ and $V_{drag}^{AS}$ are $x_S\sim2 $ and $x_{AS}\sim4 $, respectively, consistent with previous reports \cite{zheng_tunable_2024}. However, in the spin-polarized regime, the power-law exponents increased to $x_S\sim3.5 $ and $x_{AS}\sim 6 $. Using the expressions from Klesse and Stern’s theory \cite{klesse_coulomb_2000}, we extracted the $K_{c-}$ of both components at 0 T and 14 T, as shown in Fig. \ref{fig:3}(g) and Fig. \ref{fig:3}(h). Remarkably, the extracted $K_{c-}$ values overlap well between the two regimes, deviating only in the tunneling-dominated regions. This analysis, repeated across three line cuts in scan window 1 and shown in supplementary Fig. S11, consistently confirms that the $K_{c-}$ at 0 T and 14 T have similar values with deviations in high subband regimes, which is expected since the power-law exponents are only predicted for single subband wires \cite{klesse_coulomb_2000}. The same TLL parameter $K_{c-}$ extracted from different power-law exponents in 0 T and 14 T experimentally verifies the theoretical prediction of different power-law dependences in spin-full and spin-polarized regimes \cite{klesse_coulomb_2000}, yielding $K_{c-}$ values for $V_{drag}^{S}$ and $V_{drag}^{AS}$ are around 1.6 and 2.4. Crucially, this prediction for reciprocal drag in the single subband limit appears to hold even in the non-reciprocal regime and in multi-channel wires.

The magnetic field dependence of power-law exponents and the extracted values of $K_{c-}$ are further confirmed in three line cuts in scan window 2, as shown in Fig. \ref{fig:4} and Fig. S12 in the supplement. Similar to Fig. \ref{fig:3}(a) and Fig. \ref{fig:3}(b), the subband stripes and drag signal sign changes near the subband boundary are also observed. One notable feature is the shift of conductance plateaus toward more negative gate voltages with increasing temperature. This behavior can be attributed to the thermal broadening of electron distribution or to the effect of the mutual Coulomb interaction \cite{Tarucha_1995}, and the subband shift has been compensated in Fig. \ref{fig:4}(h) to the base temperature wire conductance. The temperature-dependent $V_{drag}^{AS}$ at a few selected BPL gate positions is shown in Fig. \ref{fig:4}(c) and Fig. \ref{fig:4}(d) for 0 T and 14 T, respectively. In addition to the suppression of signal amplitude at higher fields, the turning temperature $T_0$ increases from $\sim 1.6K$ at 0T to $\sim 2K$ at 14 T. The higher $T_0$ can also be verified through the 2D derivative map $\frac{dV_{drag}^{AS}}{dT}$ shown in Fig. S7. According to the theory by Klesse and Stern, the crossover temperature $T^*$ is a function of four length scales: the wire separation $\overline{d}$, wire width $d$, effective Bohr radius $a_B$, and the mean (intrawire) electron distance $r=\pi/k_F$, which will also change relative value of $K_{c-}$ \cite{klesse_coulomb_2000}. In the limit of widely separated wires $\overline{d}\gg r$, $K_{c-}<1$ and the crossover temperature $T^*$ scales as $T^*\sim E_0e^{-\frac{bk_F \overline{d}}{1-K_{c-}}}$, where $b=1$ for spin-polarized, and $b=2$ for spin-unpolarized wires, and $E_0$ is of the order of Fermi energy \cite{klesse_coulomb_2000}. For small wire separation $\overline{d}\ll r$, $K_{c-}>1$ and $T^*$ is exponentially suppressed according to $T^*\sim E_0 e^{-\frac{\pi^3a_B}{r} \frac{c(k_F)}{k_F \bar{d}}}$, where $c(k_F)$ is of order one and logarithmically depends on $k_F$. In our system, using an estimated one subband density of $n_{1D}=9.03 \times 10^7 \text{m}^{-1}$ \cite{makaju_nonreciprocal_2024} and interwire distance $\overline{d}=33nm$, we obtain $k_F \overline{d}=4.7$ which falls into the case of widely separated wires. In this case, such a formula will give us a larger $T^*$ for spin-polarized wires with $K_{c-}<1$, which agrees with the observed higher turning temperature $T_0$ at 14 T in our systems. However, the unphysical value of the estimated $T^* \sim 1.9\times 10^4 K$ and the fact $K_{c-}>1$ indicate that the large interwire separation criteria limit does not apply to our closely spaced device and suggest that the observed $T_0$ might not correspond to the predicted $T^*$. 

Within the g-ology framework,  Klesse and Stern estimate $K_{c-}=\sqrt{\frac{1+U_{c-}}{1-U_{c-}}}$ and $U_{c-}=\frac{1}{2\pi v_{c-}}(-g_2+\bar{g}_2+g_1)$, where the forward-scattering terms $g_2$ and $\overline{g_2}$, and the backscattering term $g_1$, contribute oppositely to the interaction strength $U_{c-}$ \cite{voit_one-dimensional_1995, klesse_coulomb_2000}. According to these theoretical prediction, the estimated $K_{c-}$ is around 0.994 for spinfull and 0.997 for spin-polarized wires using the screening length $D=142 nm$ and wire width $d = 50 nm$ in the single subband regime of our coupled wires. This estimate does not align with the much larger values of $K_{c-}$ reported in the letter. It is worth nothing that the estimate for $g_{1}$ was made in the limit of $k_F^{-1} \geq d$, which does not apply to our electrically defined quantum wires \cite{klesse_coulomb_2000}. In addition, this model was only calculated for the reciprocal drag signal in an ideal system, and our mesoscopic system has both strong reciprocal and nonreciprocal signals, which might account for the mismatch here. Despite the discrepancy in the estimated values, this theory can still yield $K_{c-}>1$ for repulsive interactions when the wires are in very close proximity, such that the small momentum scattering couplings, within the g-ology framework \cite{voit_one-dimensional_1995}, are nearly equal. While different features have been observed to be consistent with the prediction for the small separation between wires, the inconsistency of the predicted values of $K_{c-}$ and $T^*$ stresses the importance of further theoretical work, especially in understanding the discrepancy in parameters for both reciprocal and nonreciprocal drag signals in the non-ballistic limit.


The higher power-law exponents and reduced drag magnitude in spin-polarized regimes are shown in Fig. \ref{fig:4}(e) and Fig. \ref{fig:4}(f), consistent with results from scan window 1. Compared to the symmetric component, the antisymmetric component power-law exponent exhibits oscillations as a function of drag wire density. Such oscillations are also observed in the extracted $K_{c-}$ of the antisymmetric component at 14 T while absent at 0 T, as shown in Fig. \ref{fig:4}(g). To better study the correlations between the subband filling and exponents oscillations, the drag wire conductance of a line cut at TPL = -1.095 V and 14 T is plotted along with the 14 T antisymmetric $K_{c-}$ as a function of the BPL gate voltages in Fig. \ref{fig:4}(h). Starting from 2.6, $K_{c-}$ generally increases as the drag wire electron density increases, but is accompanied by plateaus at the same position as the defects' plateaus. The $K_{c-}$ reaches a maximum value of 3.4 when the first subband spin-up $1\uparrow$ is fully populated, then decreases as the $1\downarrow$ subband begins to fill. The $K_{c-}$ reaches a local minimum of 2.6 and shows hints of an increase when the $1\downarrow$ completes fills and the $2\uparrow$ subband starts to populate. 

This nonmonotonic evolution of the TLLs parameter $K_{c-}$ as a function of subband filling in spin-polarized quantum wires can potentially be explained by the different dependence of the scattering terms within the g-ology framework introduced previously. As the density in the $1\uparrow$ subband increases, strong screening suppresses $g_2$ more rapidly than $g_1$ and $\bar{g}_2$, leading to an enhanced $U_{c-}$ and consequently $K_{c-}>1$. The subsequent decrease in $K_{c-}$ upon filling the $1\downarrow$ subband reflects the onset of additional inter-subband scattering, which enhances repulsive interactions in the antisymmetric sector. The subband-dependent oscillations have been observed in three line cuts in scan window 2 but are absent in scan window 1, possibly because of the stronger confinement of wires and enhanced scattering. These results demonstrate the tunability of collective excitations via spin-selective subband engineering and provide direct insight into nontrivial interaction effects in multi-mode Luttinger liquids.

In conclusion, we have measured Coulomb drag between closely separated quasi-1D quantum wires in spin-polarized regimes in the presence of both reciprocal and non-reciprocal signals. Spin polarization of the 1D subbands was confirmed through conductance measurements \cite{thomas_possible_1996, graham_interaction_2003}, and corresponding peak splitting in the drag signal was observed under strong parallel magnetic fields. The sign dependence of the rectified drag signal on asymmetric electron-hole transmission probabilities, as predicted by theory \cite{Levchenko_2008}, is experimentally verified. Crucially, we investigated the distinct power-law dependencies of Coulomb drag in spin-full and spin-polarized regimes for both reciprocal and non-reciprocal signals. Consistent with theoretical prediction, the same TLL parameter $K_{c-}$ extracted from the different power-law exponents of spin-full and spin-polarized wires are first reported here, while the high value of $K_{c-}$ could be explained by the small momentum scattering terms with similar amplitude for closely separated wires \cite{voit_one-dimensional_1995, klesse_coulomb_2000}. Furthermore, we observed a nonmonotonic evolution of $K_{c-}$ as a function of drag wire density for the reciprocal Coulomb drag signal in the spin-polarized regime, which correlates with the sequential filling of spin-resolved subbands. The different power-law dependence and the same extracted interaction parameter $K_{c-}$ of Coulomb drag in spin-full and spin-polarized regimes are in good agreement with theoretical predictions, supporting the theoretical framework for TLLs. However, further theoretical work is needed to fully understand the physical origins of $K_{c-}>1$, the density dependence of the scattering mechanisms, particularly in multi-subband wires and the parameters discrepancies between the reciprocal and the non-reciprocal regimes.

\begin{acknowledgments}

\textit{Acknowledgements---} This work was performed, in part, at the Center for Integrated Nanotechnologies, an Office of Science User Facility operated for the U.S. Department of Energy (DOE) Office of Science. Sandia National Laboratories is a multimission laboratory managed and operated by National Technology $\And$ Engineering Solutions of Sandia, LLC, a wholly owned subsidiary of Honeywell International, Inc., for the U.S. DOE’s National Nuclear Security Administration under contract DE-NA-0003525. The views expressed in the article do not necessarily represent the views of the U.S. DOE or the United States Government. Part of this work was conducted at the Research Service Centers of the Herbert Wertheim College of Engineering at the University of Florida. A portion of this work was also performed at the National High Magnetic Field Laboratory. This work was partially supported by the National High Magnetic Field Laboratory through the NHMFL User Collaboration Grants Program (UCGP). The National High Magnetic Field Laboratory is supported by the National Science Foundation through NSF/DMR-1644779, NSF/DMR-2128556 and the State of Florida. A.L. acknowledges financial support by the National Science Foundation Grant No. DMR-2452658 and H. I. Romnes Faculty Fellowship provided by the University of Wisconsin-Madison Office of the Vice Chancellor for Research and Graduate Education with funding from the Wisconsin Alumni Research Foundation. \\

Data availability:\\
The data that support the findings of this article are openly available .”
\end{acknowledgments}

\bibliography{} 

\newpage
\section{End Matter}

\noindent Appendix A: Device fabrication and operations\\
The vertically integrated quantum wire device is fabricated from an n-doped GaAs/AlGaAs electron bilayer heterostructure with two 18-nm-wide quantum wells separated by a 15-nm-wide Al$_{0.3}$Ga$_{0.7}$As barrier, resulting in an interwire separation $d_{\text{vert}} = 33$ nm. The unpatterned density and mobility of the GaAs quantum well are $n = 2.98 \times 10^{11} \, \text{cm}^{-2}$ and $\mu = 7.4 \times 10^4 \, \text{cm}^2 / \text{V} \cdot \text{s}$, respectively. As shown in Fig. \ref{fig:1}(a), each wire is defined by a pinch-off (PO) gate and a plunger (PL) gate with the top and bottom gates separated by $\sim 250$ nm. The measurements were performed using standard low-frequency AC techniques at a frequency of 13 Hz in a dilution refrigerator at a base lattice temperature below 7 mK and electronic temperature of $\sim$ 15 mK.  Consistent with previous works in vertically-coupled quantum wires \cite{laroche_positive_2011, laroche_1d-1d_2014}, the PO gates are primarily used to independently contact the quantum wires and minimize tunneling current between them, while the PL gates are used to adjust the wire’s width and electronic density. With appropriate negative voltages applied to four gates, two independently contacted quantum wires are created, as shown in Fig. \ref{fig:1}(b). In this vertically superimposed design, interlayer interactions occur only in the region where the two quasi-1D wires overlap. As depicted in Fig. \ref{fig:1}(b), the drive current ($I_{\text{drive}}$) is applied to the top wire, and the induced Coulomb drag voltage ($V_{\text{drag}}$) is measured in the bottom wire. In the subsequent measurement, the bottom wire is used as the drag wire since the bottom wire exhibits sharper subbands and fewer defects compared to the top wire \cite{supp}. Additional details regarding the device characterization and consistency tests of Coulomb drag are presented in a prior publication \cite{zheng_tunable_2024} and in the supplement \cite{supp}.\\

\noindent Appendix B: Determination of tunneling regions\\
The tunneling regions shown in Fig. \ref{fig:3} are determined from both the drag wire conductance 2D scan at different magnetic fields and the out-of-phase components in the drag signal 2D maps. As shown in Fig. S10 in the supplement, the drag (bottom) wire pinches off at BPL = -1.05 V and BPL = -1.01 V in 0 T and 14 T, respectively. When the BPL gate voltage is more negative than the pinch-off value, the drag wire resistance is larger than $240\text{ k}\Omega$, and electrons prefer to tunnel across the wires, which leads to a large out-of-phase drag signal. \\

%



\end{document}